\documentstyle[pre,aps,epsfig,multicol]{revtex}
\begin{document}
\draft
\widetext

\newcommand{\be}{\begin{equation}}
\newcommand{\ee}{\end{equation}}
\newcommand{\ber}{\begin{eqnarray}}
\newcommand{\eer}{\end{eqnarray}}


\title{Dynamic States of a Continuum Traffic Equation with On-Ramp}

\author{H. Y. Lee$^{(1,2)}$, H.-W. Lee$^{(1)}$, and D. Kim$^{(1,2)}$}
\address{
$^{(1)}$Center for Theoretical Physics, Seoul National University, 
Seoul 151-742, Korea \\
$^{(2)}$Department of Physics,
Seoul National University, 
Seoul 151-742, Korea }

\maketitle

\begin{abstract}
We study the phase diagram of the continuum traffic flow model of
a highway with an on-ramp. 
Using an open boundary condition,
traffic states and metastabilities are investigated 
numerically
for several representative values of the 
upstream boundary flux $f_{\rm up}$ and for the whole
range of the on-ramp flux $f_{\rm rmp}$.
An inhomogeneous but time-independent traffic state
(standing localized cluster state) is found 
and related
to a recently measured traffic state.
Due to the density gradient near the on-ramp,
a novel traffic jam can occur even when 
the downstream density is below the critical
density of the usual traffic jam formation in homogeneous
highways, and its structure varies 
qualitatively with $f_{\rm rmp}$.
The free flow, the recurring hump (RH) state, and the traffic jam
can all coexist in a certain metastable region where the free
flow can undergo phase transitions either to the RH state
or to the traffic jam state.
We also find two nontrivial analytic solutions.
These solutions correspond to the standing localized cluster
state and the homogeneous congested traffic state
(one form of the novel traffic jam), which
are observed in numerical simulations.
\end{abstract}
\widetext


\begin{multicols}{2}

\narrowtext

\section{INTRODUCTION}
Traffic flow, a many body system of strongly interacting vehicles,
shows various complex behaviors.
Numerous empirical data of the highway
traffic have been obtained\cite{Treiterer,Kernerl,Kerner1,Kerner2,Helbing0},
which demonstrate the existence of 
distinct dynamic states and dynamic phase transitions
between them.
Recent studies reveal physical phenomena such as hysteresis,
self-organized criticality, and phase transitions in 
the traffic flow \cite{proceeding96,proceeding98}.

The transition from the homogeneous free flow to the jammed state has been 
studied by microscopic and macroscopic models without any 
inhomogeneity in the system
\cite{Bando,Nakanishi,KNagel,Schreckenberg,Prigogine,Helbing1,Kerner5}.
The traffic jam, one of the dynamic phases of the traffic flow,
appears spontaneously when the vehicle density is between the two critical
values $\rho_{\rm c1}$ and $\rho_{\rm c2}$ $(> \rho_{\rm c1})$.
The traffic jam, however, can appear even below $\rho_{\rm c1}$.
The traffic jam can be triggered by localized perturbations
provided that the density is larger than a different critical
value $\rho_{\rm b}$ ($< \rho_{\rm c1}$).
As a result, in the density range between $\rho_{\rm b}$ and $\rho_{\rm c1}$,
both the free flow and the traffic jam can exist,
resulting in metastability and hysteresis\cite{Kerner4,Kerner95,Krauss}.
It is observed that some features of the traffic jam
are uniquely determined by underlying dynamics, and 
independent of initial conditions of the traffic flow that lead to
the jam\cite{Kerner2}.
The presence of such characteristic features is also
reproduced by 
analytic and numerical studies of traffic flow models\cite{Kerner4,Kerner3}.

 The synchronized traffic flow, another dynamic phase of the traffic flow,
is identified in recent measurements on highways \cite{Kernerl,Kerner1}.
The synchronized traffic flow resembles the traffic
jam in the sense that both states produce inhomogeneous density
and flow profiles.
The dynamics of the synchronized flow is however
much more complicated than that of the traffic jam.
One notable property of the synchronized traffic flow
is the high level of its average flow, which almost matches
the flow of the free flow state.
The synchronized traffic flow is observed, in nearly all 
occasions, localized near ramps and it is thus believed that
ramps are important for the stability of the synchronized traffic flow.
The discontinuous transition from the free flow to the synchronized flow
can be induced by localized perturbations of finite amplitudes.
Measurements show a hysteresis effect in
the phase transitions between the free flow and the synchronized flow:
the transition from the synchronized flow to the free flow
occurs at a lower on-ramp flux, or lower upstream flux, 
than that for the reverse transition.
In Ref.~\cite{Lee}, the recurring hump (RH) state is proposed as an origin of 
the (nonstationary type) synchronized traffic flow \cite{Kerner1},
and the dynamic phase transitions between the RH state and the free flow 
are investigated 
using continuum traffic equations that take into account 
the effect of ramps.
In the RH state, the vehicle density and the velocity 
show temporal oscillations 
which are localized near on-ramps.
That the synchronized flow is maintained for several hours
can be explained from one important property of the RH state, its being 
a limit cycle of the traffic equations.
The RH state can be characterized as a self-excited oscillator,
where constant vehicle flux from an on-ramp serves as a source of 
the repeated excitation
and each excitation is subsequently relaxed within a localized region.
The traffic equations also describe the hysteresis phenomena between 
the RH state and the free flow.

 The traffic jam and the synchronized traffic flow
are distinct phases of traffic flow.
However, the distinction between the conditions 
for the appearance of the jam state and the 
synchronized flow is not clearly identified yet, both in measurements 
and in model studies.
Highway measurements analysis reports that 
almost identical initial states of the traffic flow can evolve 
to both the traffic jam and the synchronized flow \cite{Kernerl}.

 To describe the hysteretic phase transitions
between the free flow and the synchronized flow,
a different macroscopic model 
based on a gas-kinetic approach is also proposed \cite{Helbingl}.
In this model, a peak of the inflow from an on-ramp
produces a congested but homogeneous region near the on-ramp,
which spreads in the upstream direction.
This homogeneous congested traffic (HCT) state is proposed as 
an explanation for the (stationary type) synchronized 
traffic flow \cite{Kerner1}.
The subsequent study \cite{Helbingcondmat} investigated
the phase diagram of the model and identified
additional dynamic phases such as the standing localized cluster (SLC),
the triggered stop and go (TSG), and 
the oscillating congested traffic (OCT) states. 
Analytical conditions for the existence of these phases
are provided and it is suggested that the phase diagram
is universal for a class of traffic models.
The study is however restricted to the 
traffic states generated from a particular initial condition 
and thus important issues such as multistability and
hysteresis are not addressed. 

 In this paper, we investigate the phase diagram 
of the traffic flow in the presence of an on-ramp using a different
continuum model \cite{Kerner5},
which tests the idea of the universal phase diagram.
Various traffic states in Ref.~\cite{Helbingcondmat} are reproduced. 
However, the phase diagram is found to be qualitatively different.
For instance, some traffic states, which represent
distinct phases in Ref.~\cite{Helbingcondmat}, make smooth crossovers
to other traffic states without any sharp phase boundaries in between,
implying that they are different limiting behaviors of
a single dynamic phase.
The investigation is also performed for a large variety of
initial conditions using two effective search methods.
The conditions for the stable existence of the free flow, the RH state, 
and the traffic jam are examined.
In some parameter ranges, it is found that
multiple dynamic phases can remain stable with respect to
sufficiently small perturbations.
In such parameter ranges, finite perturbations may induce
transitions between those phases, resulting in metastability.
Due to the presence of the on-ramp,
the evolution process of the jam shows several different patterns
and the phase boundaries for the formation of the jam
are significantly modified.

 The paper is organized as follows.
In the next section, 
we investigate the possible traffic phases for given values of
the upstream flux and
the input flux through the on-ramp.
Various new features are discovered, which are absent
in homogeneous highways.
We examine the conditions for the stabilities of the traffic jam and the 
RH state. 
The metastability among the free flow, the RH state, and the traffic jam
is investigated, and
the travel time distributions of the three states
are compared.
We also discuss the several different evolution processes of the jam
due to the presence of the on-ramp.
Based on the phase diagrams,
we find that the on-ramp flux becomes a more important factor
for the formation of the traffic jam
than the total flux, the sum of the on-ramp flux and the upstream flux.
In Sec. III, 
we demonstrate analytically that our macroscopic model
possesses nontrivial solutions which are indeed found
in numerical simulations.
Finally, Sec. IV summarizes our results.

\section{Phase Diagrams of Traffic equations with an On-ramp}
In this work, we adopt the continuum model of the highway traffic
flow proposed by Kerner and Konh\"{a}user \cite{Kerner5},
\begin{eqnarray}
{\partial \rho \over \partial t}+{\partial (\rho v) \over \partial x}
&=&q_{\rm in}(t)\varphi (x) \ , \label{eq:eqofmotion1} \\ 
\rho \left({\partial v \over \partial t}+v{\partial v \over \partial x}\right)
&=&{\rho \over \tau}[V(\rho)-v] -c^2_0 \ {\partial \rho \over \partial x}
+\mu {\partial^2 v \over \partial x^2} \ ,
\label{eq:eqofmotion2}
\end{eqnarray}
where $\rho(x,t)$ is the local vehicle density and $v(x,t)$ the local
velocity. $q_{\rm in}(t)\varphi (x)$ is the
source term  representing the external flux through an on-ramp. 
The spatial distribution of the external flux $\varphi (x)$
is localized near $x=0$ (on-ramp position) and 
normalized so that $q_{\rm in}(t)$ denotes 
the total incoming flux.
$V(\rho)$ is the safe velocity that  
is achieved in the time-independent and homogeneous traffic flow.
In Eq.~(\ref{eq:eqofmotion2}), the second term on the right hand
side represents an effective ``pressure'' gradient
on vehicles due to the anticipation driving \cite{Kerner5}
and the velocity fluctuations \cite{Helbing0,Helbing1}, 
and the third term takes into account an intrinsic 
dampening effect that is required to
fit the experimental data \cite{Kuhne}.
Here $\tau,c_0, \mu$ are appropriate constants.
The flux or flow, $\rho v$, is denoted below by either $q$ or $f$.

 In order to investigate the effects of a single on-ramp,
we use the open boundary condition.
The upstream boundary values of the density and velocity are fixed
at $\rho(x=-L/2,t)=\rho_{\rm up}$ and $v(x=-L/2,t)=V(\rho_{\rm up})$,
respectively. On the other hand, the values 
at the downstream boundary ($x=L/2$) 
are linearly extrapolated from their values 
at neighboring points, $x=L/2-\Delta x$ and $L/2-2\Delta x$
where $\Delta x$ is spacing used in the discretization.
The numerical simulations are performed 
using the two-step Lax-Wendroff scheme\cite{Press}.
We choose the following parameters : 
$\tau=$ 0.5 min, $\mu=$ 600 vehicles km/h, 
$c_0=54$ km/h, and
$V(\rho)=V_0  (1-\rho/\hat{\rho})/( 1+E (\rho/ \hat{\rho})^4 )$ 
where the maximum density $\hat{\rho}=$ 140 vehicles/km, 
$V_0=120$ km/h, and $E=100$ \cite{commentV}. 
Concerning the discretization, spatial intervals of 
$\Delta x=37.8$ m and time intervals of
$\Delta t= 10^{-4}$ min are used.
We choose the spatial distribution of the external flux as
$\varphi(x)=(2 \pi \sigma^2)^{-1/2} \exp{(-x^2/{2 \sigma^2})}$
with $\sigma=56.7$ m.
With this choice of parameters, critical values are 
$\rho_{\rm b}=21.1$ vehicles/km,  
$\rho_{\rm c1}$=25.3 vehicles/km,
$\rho_{\rm c2}$=62.3 vehicles/km,
$f_{\rm b}
\equiv \rho_{\rm {b}} V(\rho_{\rm {b}})=2047$ vehicles/h, $f_{\rm c1}
\equiv \rho_{\rm {c1}} V(\rho_{\rm {c1}})=2249$ vehicles/h,
and $f_{\rm c2} \equiv \rho_{\rm {c2}} V(\rho_{\rm {c2}})=843$ vehicles/h.
The maximum flow that can be achieved in the time-independent
homogeneous flow is
$f_{\rm max}=\mbox{max}_{\rho} \{ \rho V(\rho) \}=2336$ vehicles/h.
Below we are interested mainly in the low density regime ($ < \rho_{\rm c1}$),
and thus we will use for brevity the subscript c in place of c1.

 In the real highway traffic,
there are many kinds of noises which perturb
the traffic out of its steady states.
The real traffic state is hence under an infinite sequence of
perturbations and subsequent responses of dynamic states.
Previous studies \cite{Kerner4,Kerner3} on highway traffic without ramps
however showed that 
many observed features of the traffic flow can be
explained from the steady state properties of the continuum
model without any noise.
Motivated by previous successes, we
will ignore noises in this paper.

 In the absence of noises, each dynamic phase of the traffic flow
corresponds to a steady state, or equivalently an attractor
of the nonlinear hydrodynamic model 
[Eqs.~(\ref{eq:eqofmotion1},\ref{eq:eqofmotion2})].
A steady state may exhibit complicated time dependences 
depending on the nature of the corresponding attractor.
We examine in this Section the linearly stable steady states (or phases
of traffic flow) for a given upstream flux $f(-L/2)=f_{\rm up} \equiv 
\rho_{\rm up}V(\rho_{\rm up})$ and the vehicle flux $q_{\rm in}(t)=f_{\rm rmp}$
through the on-ramp at $x=0$. 
Linearly stable states are, however, often unstable to large perturbations
and multistability can occur.
Here it is worth emphasizing that the concept of the multistability in
dynamic systems is somewhat different from that in equilibrium systems.
In equilibrium systems the free energy 
selects one particular state as a ``true'' stable
state and other states become metastable.
In dynamic systems, on the other hand, the free energy cannot
be defined and the concept of the ``true'' stable state
is not applicable. In this sense, all states
are metastable and they all should be treated equally.

 Possible presence of multiple steady states makes it very difficult
to search completely for all phases that are stable for given parameters,
since it requires examinations of many different initial conditions.
To search out all multiple steady states,
we use two methods : One is to apply
a triggering pulse to a steady state, for example, by changing the value
of $f_{\rm rmp}$ for a short time.
For a sufficiently strong pulse, a transition to a different steady state
can be induced, allowing the identification of a new steady state.
The other is the adiabatic sweeping method.
Starting with a given steady state for a particular set of the
system parameters, 
one increases or decreases one parameter adiabatically.
This way, one can find the range of the parameter
values where a dynamic state remains stable.
These two methods effectively simulate a large variety
of initial conditions.
Using these, we investigate the steady
states for given system parameters $f_{\rm up}$ and $f_{\rm rmp}$.
In particular, we concentrate on three representative values
of $f_{\rm up}$, and for each of them construct
a phase diagram for the entire range of $f_{\rm rmp}$.
However, since too large a value of $f_{\rm rmp}$
is unrealistic, we restrict our attention to the range 
$f_{\rm rmp} \leq f_{\rm rmp}^{\rm max} \equiv f_{\rm max}-f_{\rm up}$.

 The values of $f_{\rm up}$ studied in this work are
chosen from the following considerations.
In the previous studies of homogeneous highways
without ramps, it was found that the  
flux $f_{\rm b}$
provides an important boundary.
Whereas the free flow is the only stable phase
below $f_{\rm b}$,
the traffic jam can be created above this value.
In the presence of ramps, 
one can expect that 
the appearance of the traffic jam depends on whether the
upstream flux $f_{\rm up}$ is larger or smaller than $f_{\rm b}$.
(Below we show that this expectation is not true,
due to the nontrivial effect of an on-ramp.)
This property motivated us to choose one representative 
value of $f_{\rm up}$ in the range larger than $f_{\rm b}$
and another smaller.
We also choose a very small value for $f_{\rm up} (\ll f_b)$,
which later reveals the importance of $f_{\rm rmp}$ on the formation of
the congested traffic.

\subsection{$f_{\rm up} > f_{\rm b}$}
 The phase diagram of the traffic states for 
$f_{\rm up}=2119$ vehicles/h appears in Fig.~\ref{fig:phase}(a).
Here $f_{\rm rmp}^{\rm c}$ is the critical input flux through the on-ramp
above which the free flow in the downstream of the on-ramp becomes
linearly unstable.
The critical on-ramp flux $f_{\rm rmp}^{\rm c}$ is determined from
\begin{equation}
f_{\rm rmp}^{\rm c}=f_{\rm c}-f_{\rm up} \ ,
\end{equation}
where $f_{\rm c}=f_{\rm c1}$.
For $0 \leq f_{\rm rmp} \leq f_{\rm rmp}^{\rm c}$,
the flux, both in the upstream and downstream,
is lower than $f_{\rm c}$ but higher than $f_{\rm b}$.
Hence the traffic jam can be created from the free flow
by triggering events but it does not appear spontaneously.
The finite amplitude perturbation to $f_{\rm rmp}$
generates a cluster, which grows to a traffic jam
since the upstream flux is larger than $f_{\rm b}$.
The traffic jam propagates to the upstream with its characteristic
group velocity.

When $f_{\rm rmp}>f_{\rm rmp}^{\rm c}$, 
the flux of the free flow in the downstream 
is larger than the critical flux $f_{\rm c}$ and 
the free flow is linearly unstable with respect to 
long wavelength perturbations of infinitesimal amplitude.
The growth of infinitesimal perturbations leads
to spontaneously formed clusters and as pointed out in Ref.\cite{Kerner95},
complex sequences of traffic jams may appear in the downstream region.
In a certain range of $f_{\rm rmp}$,
we also observe that clusters form a periodic regular sequence.
It turns out that this regular sequence is caused by the
presence of the on-ramp, whose detailed discussion will
be given in the next subsection.

\subsection{$f_{\rm up} < f_{\rm b}$}
 In Fig.~\ref{fig:phase}(b), we present the phase diagram 
for $f_{\rm up}=1948$ vehicles/h.
The free flow can exist until $f_{\rm rmp}$ reaches $f_{\rm rmp}^{\rm c}$,
as in the previous subsection.
When $f_{\rm rmp}$ is smaller than 92 vehicles/h,
the free flow (with a transition layer) is the only stable phase.

 For $f_{\rm rmp} > 92$ vehicles/h, we find
another time-independent state beside the free flow,
which is shown in Fig.~\ref{fig:SLC}(a).
In our simulation, this new state can be generated from the free flow
by applying the triggering pulse in $f_{\rm rmp}$
for a short time.
Far away from the on-ramp, the density and flow
are homogeneous both in the upstream and downstream.
Near the on-ramp, a localized cluster appears, which does not
propagate in either direction but stays motionless.
Due to this immobility, such state is named as the
``standing localized cluster" (SLC) state in Ref.\cite{Helbingcondmat}.
The immobility of the SLC state is in contrast to the situations without
ramps where all inhomogeneities should propagate.
Hence the property is due to a novel effect of the on-ramp. 
Another interesting property of the SLC state
becomes manifest in the density-flow relations.
Notice that the density-flow relations [circles in Fig.~\ref{fig:SLC}(b)]
measured at several locations near the on-ramp do not necessarily fall on 
the homogeneous density-flow
relation curve ($\rho, \rho V(\rho)$) [solid line in Fig.~\ref{fig:SLC}(b)]
even though the relation at each measurement location 
remains stationary with time.
More remarkably, the circles lie in the linearly
unstable density region.

Incidentally, an experimental data which may be relevant
to this has been reported \cite{Kerner1}.
It was observed that when the traffic is in the stationary 
synchronized flow state,
the density and flux can remain stationary during
a relatively long time interval (2-5 min).
Their stationary values often lie 
in the linearly unstable density region
and they form a two-dimensional area in the density-flow plane
instead of falling on a single well-defined density-flow relation curve.
In Ref.\cite{Kerner1}, the stationary values are interpreted
as an indication of the spatially {\it homogeneous} traffic, and
Helbing, Hennecke, and Treiber \cite{Helbingcondmat} proposed 
the HCT state as an origin of the stationary synchronized flow.
The HCT state provides an explanation for the stability of the traffic
in the linearly unstable density region but it leads
to the formation of the well-defined density-flow relation
curve, failing to explain the absence of such a curve
in the measurement.

Present analysis of the SLC state 
raises an alternative possibility.
The SLC state shows that being stationary does not 
necessarily imply the homogeneity, and
it also explains the stability in the linearly unstable density region. 
Furthermore it can explain the absence of the well-defined
density-flow relation curve. 
We mention that upon the adiabatic variations of 
$f_{\rm up}$, $f_{\rm rmp}$ and the external flux profile $\varphi(x)$,
the density-flow relation at a single measurement location
can cover a two-dimensional area in the density-flow plane.
These agreements raises an interesting possibility
of an alternative explanation for the stationary
synchronized traffic flow based on the SLC state.
We judge however that it is yet premature to draw
a definite conclusion from these agreements alone.
Further experimental investigation of the stationary synchronized
traffic flow is necessary.
In the next Section, we demonstrate analytically that
the traffic equations (\ref{eq:eqofmotion1},\ref{eq:eqofmotion2})
do have the SLC state solution.

 As the on-ramp flux $f_{\rm rmp}$ increases adiabatically, one finds
the phase transition from the SLC state to the
recurring hump (RH) state [Fig.~\ref{fig:RH}(a)].
In the RH state, a cluster, or a hump, does not remain stationary
but moves back and forth in a localized region near the on-ramp.
Its drift to far upstream is not
allowed since the upstream vehicle density is lower than
the boundary value $\rho_{\rm b}$.
The RH state is investigated in detail in Ref.~\cite{Lee}
using the periodic boundary condition,
and many interesting properties are found such as the
discontinuous transition from the free flow to the RH state
induced by localized perturbations of finite amplitudes,
hysteresis,
gradual spatial transitions from the RH
state to the free flow, and synchronized oscillations.
These properties are identical to those of the
synchronized flow (nonstationary type) \cite{Kernerl,Kerner1},
and based on these common properties, the RH state is
proposed as the origin of the synchronized flow.

 In addition to the properties of the RH state discussed
in Ref.~\cite{Lee}, we investigate here the transition between 
the RH state and the SLC state.
Our simulation shows that the transition from the SLC state
to the RH state and the reverse transition occur at the same
critical value of $f_{\rm rmp}$ 
without hysteresis.
We also examine the oscillation amplitude of the RH state.
The amplitude decreases to zero continuously as $f_{\rm rmp}$
approaches the critical value [Fig. \ref{fig:RH}(b)].
Below the critical value, the hump does not oscillate and it
becomes a standing cluster.
These properties suggest that these transitions are a result 
of the supercritical (or very weak subcritical) 
Hopf bifurcation \cite{Jackson} of the SLC state to the RH state.

 These transitions between the SLC state and the 
RH state are not observed in the previous study \cite{Lee}, where
the adiabatic decrease of the ramp flux 
leads to the discontinuous transition of the RH state to the free flow
instead (Fig.~3 in Ref.~\cite{Lee}).
We attribute this difference to the different boundary condition adopted in
this paper.
Unlike the open boundary condition where $f_{\rm up}$ and
$f_{\rm rmp}$ can be controlled independently, 
the periodic boundary condition used in \cite{Lee}
is such that the increase (decrease) of $f_{\rm rmp}$ is always
accompanied by the decrease (increase) of $f_{\rm up}$
since the average density of the total system is fixed. 
Therefore the ``scanning" direction in Ref.\cite{Lee} 
is different from that in this paper. 

 We next discuss the traffic jam state.
In homogeneous highways without ramps, the formation
and propagation of the jam can not occur when the
flux is smaller than $f_{\rm b}$.
In the present case with an on-ramp,
the flux $f_{\rm up}$ in the upstream region is lower than $f_{\rm b}$
while the flux in the downstream region 
can be controlled by $f_{\rm rmp}$.
Thus a usual jam that consists of a single localized cluster
should decay after they reach the upstream region. So in this
sense, a usual traffic jam is not a steady state.
Our investigations show that
a different type of traffic jams (Fig.~\ref{fig:OCT})
can occur even when $f_{\rm up} < f_{\rm b}$ 
due to the nontrivial effect of the on-ramp:
Clusters are self-generated near the on-ramp repeatedly,
forming a ``train'' of clusters moving upstream.
Although each constituting cluster decays during its upstream
propagation, the train can still remain stable
provided that the decay rate is smaller than the self-generation
rate, which is controlled by
the extent of the inhomogeneity, $f_{\rm rmp}$, 
rather than by the upstream or downstream flux.
Thus the stability limits of the new traffic jam
do not coincide with those of the usual traffic jams 
[Fig.~\ref{fig:phase}(b)].
Notice that this train structure 
is different from usual traffic jams in homogeneous highways.
To indicate the structural difference, 
we will call this state the ``oscillating congested traffic" (OCT) state.

 We note that a traffic jam state very similar to the OCT state
appears for $f_{\rm up} >f_{\rm b}$.
Clusters are self-generated near the on-ramp
repeatedly, forming a regular sequence of clusters.
For $f_{\rm up}>f_{\rm b}$, however, 
each cluster does not decay in their upstream 
movement because of the high upstream density. 

 The structure of the OCT state can be compared to the RH state.
In both states, clusters appear recurrently near the on-ramp.
In the OCT state, however, the area of the congested
region expands with time, while in the RH state, clusters are localized.
This difference is due to the larger size of clusters in the former.
  
 It is also worth mentioning that the structure of the OCT state
shows an interesting crossover as $f_{\rm rmp}$ varies.
For small values of $f_{\rm rmp}$ (close to the lower stability limit),
the distance between the clusters is relatively large so that there exist
homogeneous flow regions in between [Fig.~\ref{fig:OCT}(a)].
As $f_{\rm rmp}$ increases, the distance between the clusters shrinks
and for sufficiently large values of $f_{\rm rmp}$,
the homogeneous regions between them disappear [Fig.~\ref{fig:OCT}(b)],
and the clusters are ``closely packed" inside the congested region.

 In Ref.~\cite{Helbingcondmat}, 
these structural differences are discovered using 
a different hydrodynamic model and the OCT states
for small and large $f_{\rm rmp}$
are identified as two distinct phases. The former was called 
the ``triggered stop and go'' (TSG) flow and the latter OCT.
In this paper, however, we find that these apparently different states
transform smoothly to each other as $f_{\rm rmp}$ is varied, without
any signature of singularities. Thus we group these two
states as a single dynamic phase in this paper.
This difference between this paper and Ref.~\cite{Helbingcondmat}
may be due to the different models used, but presently
we do not know the precise origin of the difference.

 We emphasize that in a certain range of 
$f_{\rm rmp}$, three phases,
the free flow, the RH state, and the traffic jam (OCT) can coexist.
In this metastable region of $f_{\rm rmp}$,
small differences in the initial traffic condition
may result in quite different final states.
We mention that in a recent measurement \cite{Kernerl},
very similar initial states of the free flow are observed to undergo
different phase transitions either to the synchronized flow
or to the traffic jam.
Here we obtain the three phases from the traffic equations
with a fixed parameter set.

 The difference between the three phases, 
the free flow, the RH state, and the traffic jam (OCT), 
is manifest in the travel time distributions which are shown
in Fig.~\ref{fig:travel time}.
In order to calculate the exact travel time distributions,
we determine the trajectory of a vehicle,
which is initially located at $x_{\rm veh}(t_0)$,
as follows:
\begin{equation}
x_{\rm veh}(t)=x_{\rm veh}(t_0)+\int_{t_0}^{t} dt' ~v(x_{\rm veh}(t'),t') \ .
\end{equation}
From the trajectory of each vehicle,
we obtain the vehicle travel time passing through the region
from $x_0=-5$ km to $x_1=5$ km.
With the same $f_{\rm up}=1948$ vehicles/h and $f_{\rm rmp}=222$ vehicles/h,
the travel time distributions of the three states show different behaviors. 
While it consists of a single peak for the free flow,
those for the RH state and the traffic jam (OCT)
show broad distributions due to the nonstationary nature of these phases.
Also notice that in average,
the travel time for the traffic jam (OCT) is greater than
those for the free flow and the RH state.

\subsection{$f_{\rm up} \ll f_{\rm b}$}
\begin{figure}
\begin{center}
\epsfig{file=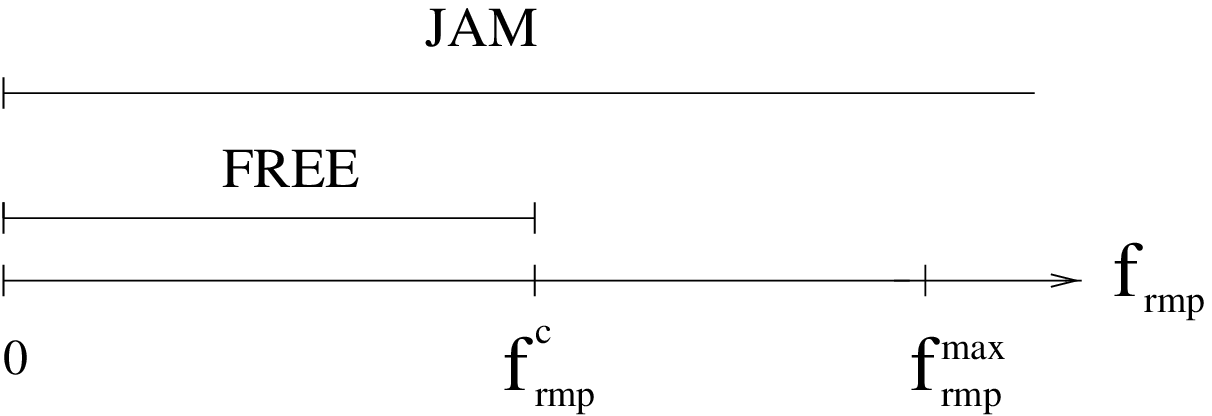,clip=,width=0.95\columnwidth}(a)\\[1cm]
\epsfig{file=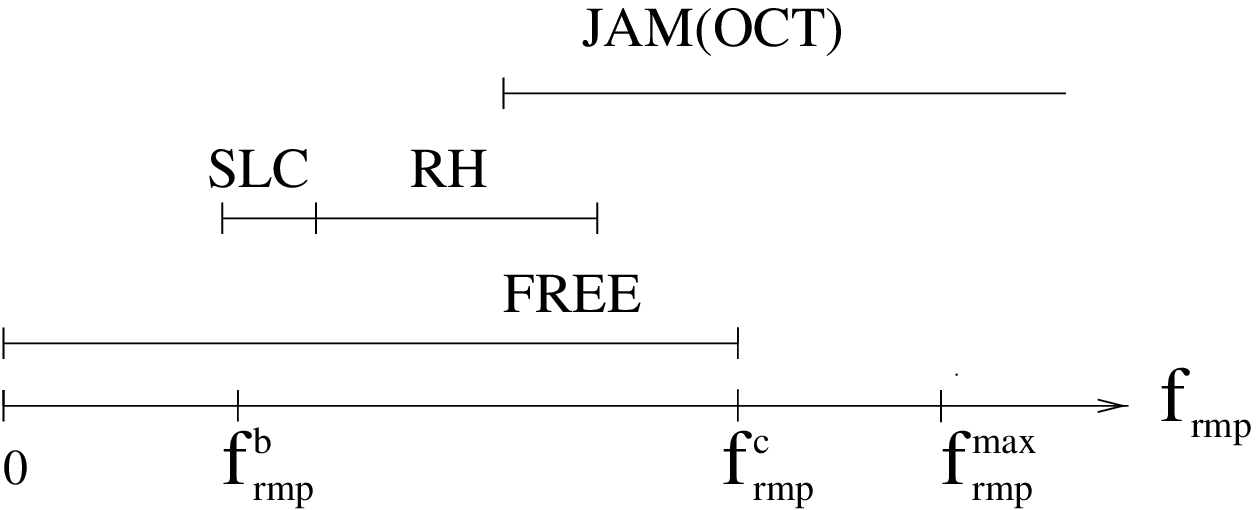,clip=,width=0.95\columnwidth}(b)\\[1cm]
\epsfig{file=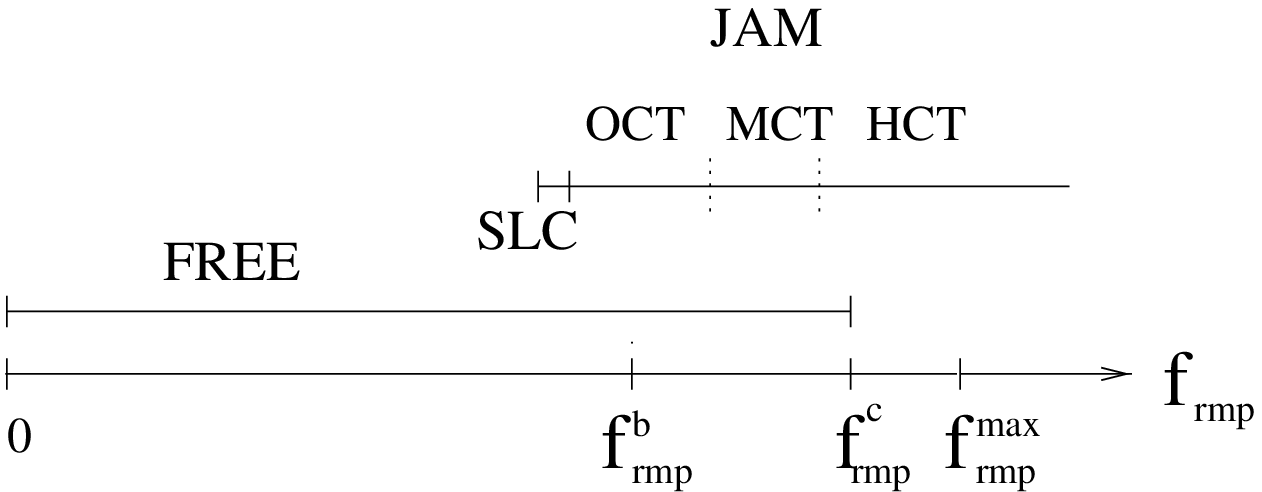,clip=,width=0.95\columnwidth}(c)
\end{center}
\caption[]{
The phase diagrams for
$f_{\rm up} > f_{\rm b}$ (a), $f_{\rm up} < f_{\rm b}$ (b), and 
$f_{\rm up} \ll f_{\rm b}$ (c).
Here $f_{\rm b}=2047$ vehicles/h, $f_{\rm c}=2249$ vehicles/h,
and $f_{\rm max}$=2336 vehicles/h
while $f_{\rm rmp}^{\rm b} \equiv f_{\rm b}-f_{\rm up}$,
$f_{\rm rmp}^{\rm c} \equiv f_{\rm c}-f_{\rm up}$,
and $f_{\rm rmp}^{\rm max} \equiv f_{\rm max}-f_{\rm up}$.
(a) The phase diagram for $f_{\rm up}=2119$ vehicles/h where
$f_{\rm rmp}^{\rm c}=130$ vehicles/h,
and $f_{\rm rmp}^{\rm max}$=217 vehicles/h.
For $f_{\rm rmp} > f_{\rm rmp}^{\rm c}$, the
traffic jam state is generated spontaneously.
(b) The phase diagram for $f_{\rm up}=1948$ vehicles/h where 
$f_{\rm rmp}^{\rm b}=99$ vehicles/h,
$f_{\rm rmp}^{\rm c}=301$ vehicles/h, and 
$f_{\rm rmp}^{\rm max}=389$ vehicles/h.
The metastable region among the free flow, the RH state,
and the OCT state extends from  $f_{\rm rmp}=206$ vehicles/h to
238 vehicles/h.
(c) The phase diagram for $f_{\rm up}=1497$ vehicles/h where
$f_{\rm rmp}^{\rm b}$=550 vehicles/h,
$f_{\rm rmp}^{\rm c}=752$ vehicles/h, and 
$f_{\rm rmp}^{\rm max}=839$ vehicles/h.
As the on-ramp flux increases, 
the traffic jams with different structures appear.
The lower stability limit
of the HCT state is defined as the value of $f_{\rm rmp}$ above
which the inhomogeneous part disappears.
In a similar way, one may define the upper stability limit of the OCT
as the value of $f_{\rm rmp}$ below which the homogeneous
part does not expand with time.
Notice that between these two stability limits, there exists
an intermediate
range of $f_{\rm rmp}$ for which both the OCT-like part (inhomogeneous part)
and the HCT-like part (homogeneous part)
expand with time.
This intermediate state is called the ``mixed congested traffic" (MCT) state.
The stability limit between the OCT and MCT state
is $f_{\rm rmp}=603$ vehicles/h
and that between the MCT and HCT state is $f_{\rm rmp}=730$ vehicles/h.
}
\label{fig:phase}
\end{figure}

\begin{figure}
\begin{center}
\epsfig{file=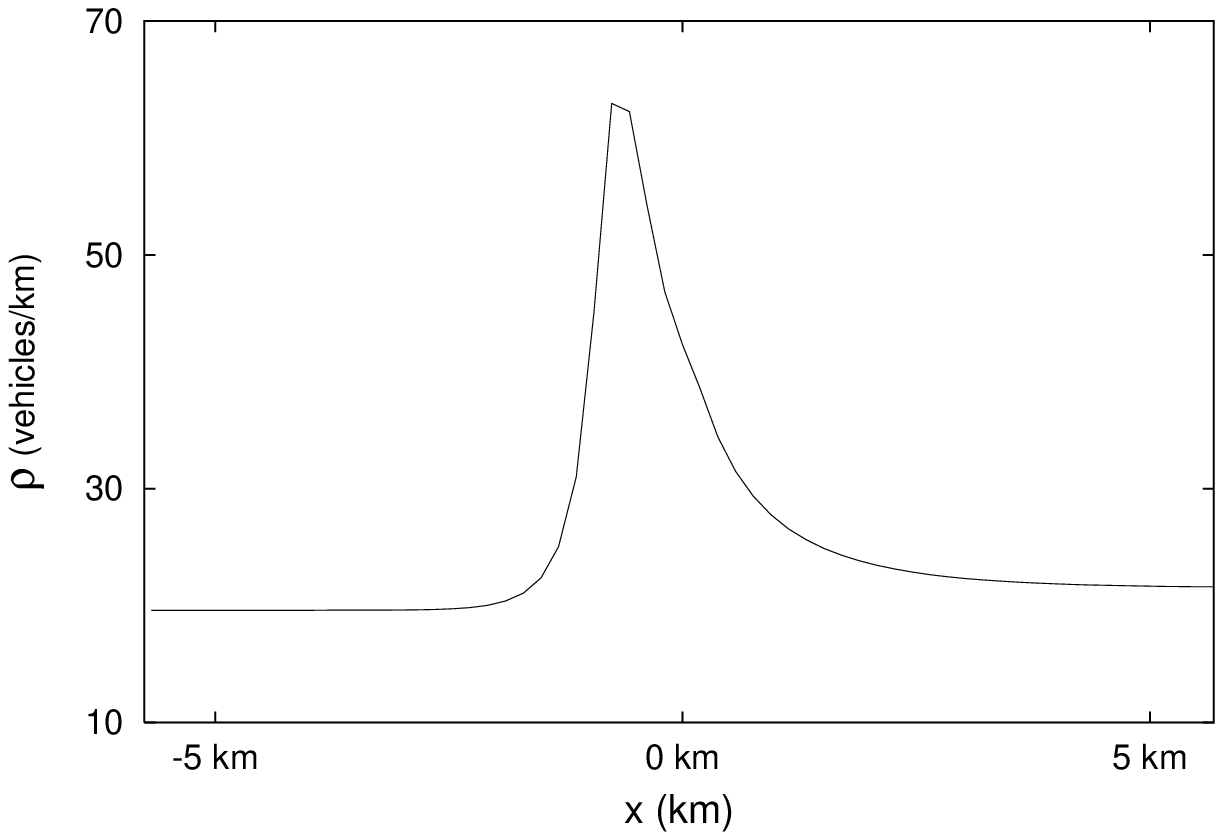,clip=,width=0.95\columnwidth}(a)\\
\epsfig{file=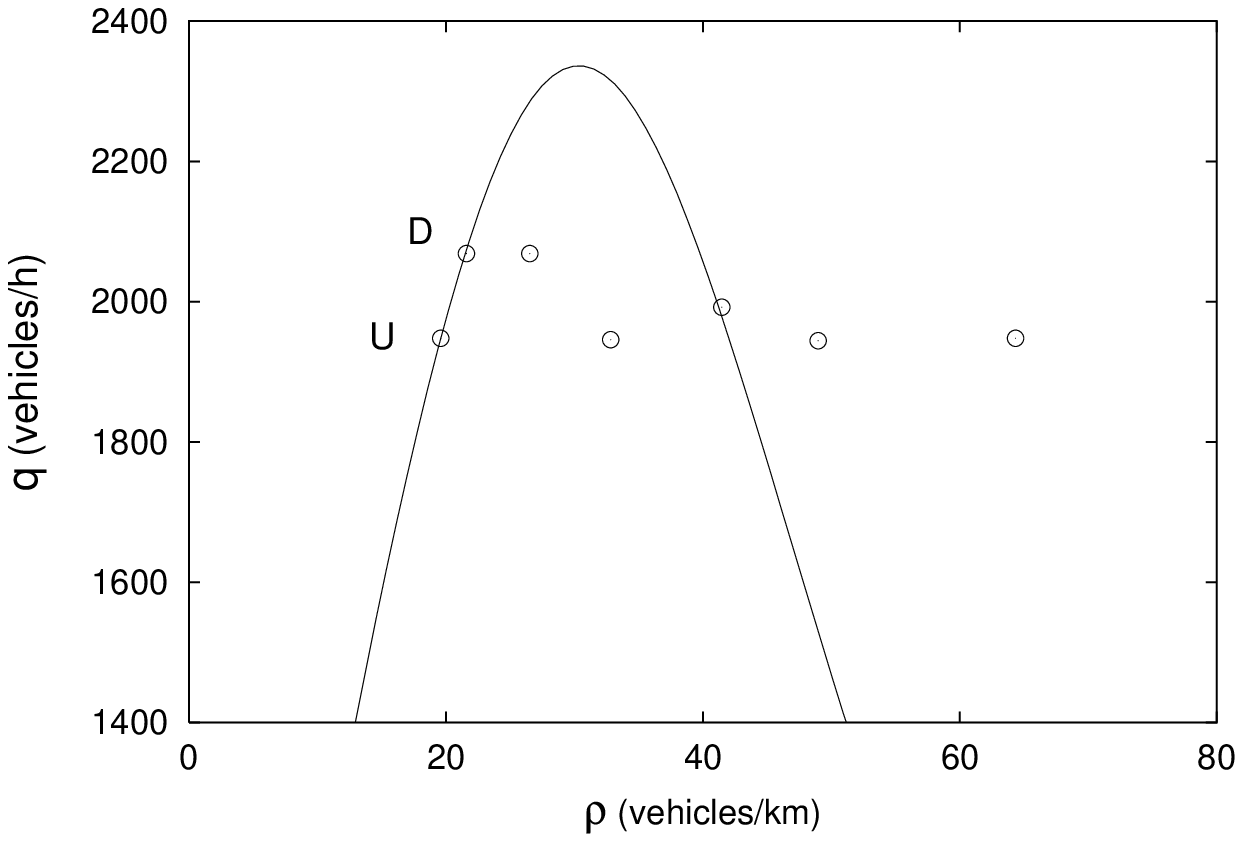,clip=,width=0.95\columnwidth}(b)
\end{center}
\caption[]{
(a) The spatial density profile of the SLC state
for $f_{\rm up}=1948$ vehicles/h and $f_{\rm rmp}=121$ vehicles/h.
The on-ramp is at 0 km.
The profile does not change with time.
(b) Circles: The density-flow relation for several positions
near the on-ramp. 
That denoted as U(D) is the data in the upstream (downstream) 
homogeneous region.  
Each data point is stationary with time, and 
upon adiabatic variations of $f_{\rm up}$ and the external
flux profile $\varphi(x)$, it covers a two-dimensional area. 
Solid line: The curve $q=\rho V(\rho)$.
}
\label{fig:SLC}
\end{figure}

\begin{figure}
\begin{center}
\epsfig{file=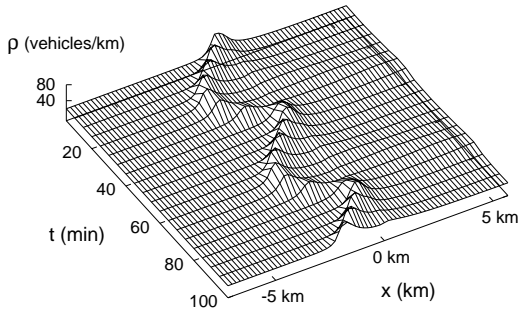,clip=,width=0.95\columnwidth}(a)\\
\epsfig{file=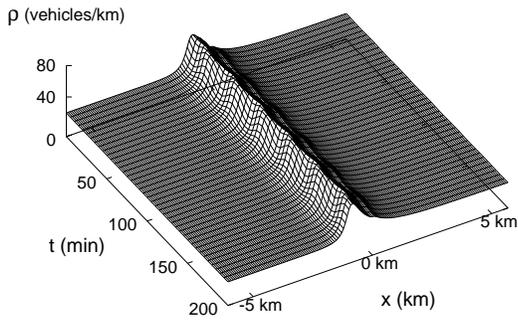,clip=,width=0.95\columnwidth}(b)
\end{center}
\caption[]{
(a) The evolution of the RH state for $f_{\rm up}=1948$ vehicles/h
and $f_{\rm rmp}=222$ vehicles/h.
The hump moves back and forth in a localized
region near the on-ramp.
(b) The evolution of the RH state for $f_{\rm up}=1948$ vehicles/h
and $f_{\rm rmp}$=130 vehicles/h.
Near the lower stability limit of the RH state,
which coincides with the upper stability limit
of the SLC state, the oscillation amplitude of the RH state is very small.
}
\label{fig:RH}
\end{figure}

\begin{figure}
\begin{center}
\epsfig{file=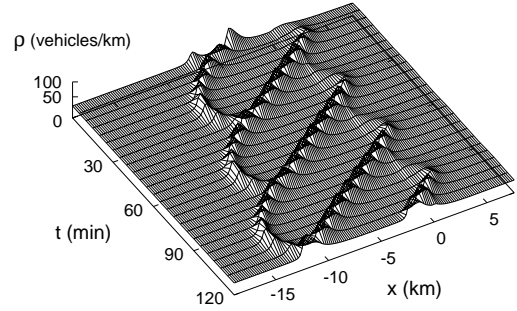,clip=,width=0.95\columnwidth}
\epsfig{file=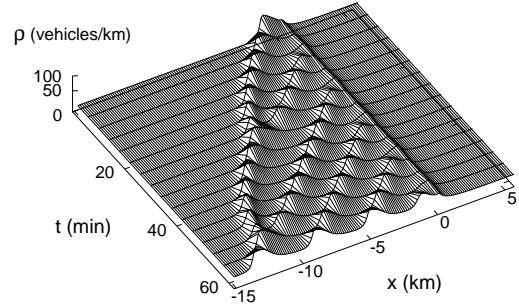,clip=,width=0.95\columnwidth}
\end{center}
\caption[]{
(a) The spatiotemporal evolution of the density of the OCT
state for $f_{\rm up}=1948$ vehicles/h and $f_{\rm rmp}=222$ vehicles/h.
The values of $f_{\rm up}$ and $f_{\rm rmp}$
are the same with those in Fig.~\ref{fig:RH}(a).
The RH state in Fig.~\ref{fig:RH}(a) and the OCT state in this figure
are independent metastable states.
(b) The spatiotemporal evolution of the OCT state
for $f_{\rm up}=1948$ vehicles/h and $f_{\rm rmp}=381$ vehicles/h.
The increase of $f_{\rm rmp}$ generates  
the ``closely packed'' clusters.
Similarly to (a), each cluster decays
at far upstream from the on-ramp.
}
\label{fig:OCT}
\end{figure}

\begin{figure}
\begin{center}
\epsfig{file=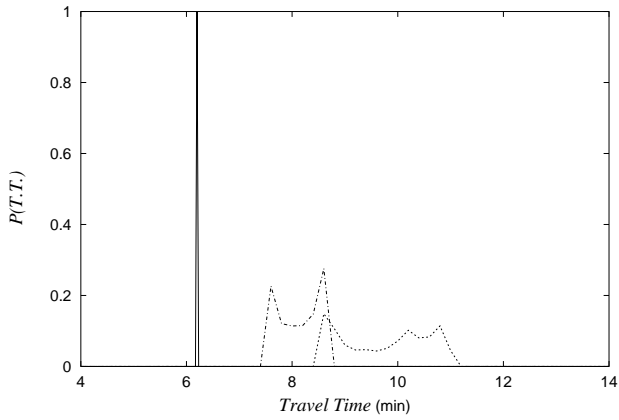,clip=,width=0.95\columnwidth}
\end{center}
\vspace{0.5cm}
\caption[]{
The travel time distributions of the free flow (solid line),
the RH state (dashed line), and the traffic jam (OCT) (dotted line)
for $f_{\rm up}=1948$ vehicles/h and $f_{\rm rmp}=222$ vehicles/h.
The travel time distributions are obtained by following  
$10^5$ trajectories of vehicles through the region
from -5 km to 5 km.
}
\label{fig:travel time}
\end{figure}

 Fig.~\ref{fig:phase}(c) shows the phase diagram 
for $f_{\rm up}=1497$ vehicles/h (about 25\% lower than $f_{\rm b}=2047$
vehicles/h).
The free flow remains linearly stable for $f_{\rm rmp} < f_{\rm rmp}^{\rm c}$.
In a narrow range of $f_{\rm rmp}$,
480 vehicles/h$\leq f_{\rm rmp} <$ 492 vehicles/h,
the SLC state is found, and for $f_{\rm rmp} \geq 492$ vehicles/h,
the OCT state is found. 
For this low upstream flux, however, the RH state does not appear.
We find the critical value of $f_{\rm up}$
below the RH state is absent is about 1872 vehicles/h. 

 It is interesting to notice that the upper stability limit of
the SLC state and the lower stability limit of the OCT
state coincide within our numerical accuracy.
We verified that 
upon the adiabatic increase of $f_{\rm rmp}$,
the SLC state undergoes the phase transition to the OCT state,
and upon the adiabatic decrease of $f_{\rm rmp}$,
the reverse phase transition occurs, both 
at $f_{\rm rmp}=492$ vehicles/h.
This coincidence raises an interesting possibility of a close
relation between the two phases.
This possibility is also supported by the expansion rate
of the congested region,
which seems to approach zero smoothly
as $f_{\rm rmp}$ is reduced to
the lower stability limit of the OCT state.
 
\begin{figure}
\begin{center}
\epsfig{file=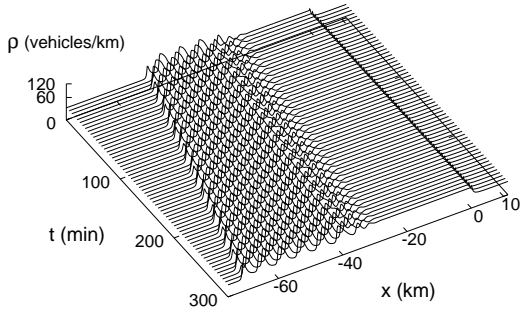,clip=,width=0.95\columnwidth}(a)\\
\epsfig{file=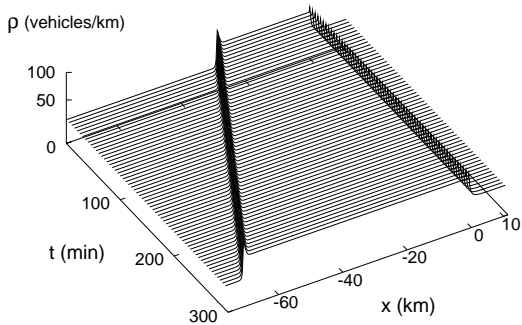,clip=,width=0.95\columnwidth}(b)
\end{center}
\caption[]{
The evolution of the MCT (a) and HCT state (b) for $f_{\rm up}$=1497 vehicles/h.
(a) $f_{\rm rmp}$=635 vehicles/h.
The congested region consists of the homogeneous part and
the inhomogeneous part.
(b) $f_{\rm rmp}$=794 vehicles/h.
The inhomogeneous part is not present.
The upstream front  
moves with a fixed group velocity.
}
\label{fig:OCTstructure}
\end{figure}

We next examine the evolution of the traffic jam state.
Fig.~\ref{fig:OCTstructure} shows the evolution of the
structure of the congested region as $f_{\rm rmp}$
is increased. For relatively small $f_{\rm rmp}$,
the structure is the same as in Fig.~\ref{fig:OCT}.
As $f_{\rm rmp}$ increases, however, a homogeneous flow region
appears near the on-ramp,
which expands with time [Fig.~\ref{fig:OCTstructure}(a)].
Hence the congested region is partitioned into 
an inhomogeneous part and a homogeneous one.
For an even higher value of $f_{\rm rmp}$ (=730 vehicles/h),
the inhomogeneous part shrinks in length with time and after
this transient process, the whole congested region consists
of a homogeneous part [Fig.~\ref{fig:OCTstructure}(b)].
In Ref.\cite{Helbingcondmat}, this state of traffic flow is
named the ``homogeneous congested traffic" (HCT) state
and is identified as a distinct phase.

Unlike Ref.~\cite{Helbingcondmat}, however, it is not
so clear in our simulations whether the OCT and HCT states are distinct
phases.
The distinction between the OCT and the HCT state is obscured further in
our simulations by the presence of an intermediate traffic state
where both the OCT-like inhomogeneous part and the HCT-like homogeneous part
expand with time. We call this intermediate state the
``mixed congested traffic" (MCT) state [Fig.~\ref{fig:OCTstructure}(a)].
As $f_{\rm rmp}$ increases, the change from the OCT state to
the MCT state and then to the HCT state
seems to occur in a smooth way.
We thus infer that the OCT, MCT, and HCT are different forms
of a single jam phase.
 
 We now focus on the HCT state.
Notice that even though $f_{\rm up}< f_{\rm b}$, 
the area of the congested traffic increases 
monotonically with the group velocity of the upstream front
about $-7.68$ km/h, which is considerably lower than the usual
jam propagation velocity $\sim-$15 km/h \cite{Kerner2}. 
This monotonic widening of the congested region
is caused by the ``blockage'' effect of the ramp,
which can be understood easily by recalling
that in real highways, a large flux from the on-ramp
can almost block the flow of vehicles on main highways.
We also mention that the structure of the HCT state is identical
to the traffic jam (shock) caused by a blockage \cite{Janowsky},
which implies that for large $f_{\rm rmp}$,
the ramp works as a bottleneck.

Interestingly, the density in the congested region lies 
in the linearly unstable region of the homogeneous flow, 
$\rho_{\rm c1} (=\rho_{\rm c})< \rho < \rho_{\rm c2}$. 
Hence according to Refs.~\cite{Kerner5,Kerner4},
long wavelength fluctuations of even infinitesimal amplitude should
grow in this region.
In our simulations, we find that small inhomogeneities in the initial
state indeed grow to form clusters.
These clusters however disappear when they reach the upstream
boundary of the congested region and the congested region
becomes homogeneous afterwards.
This result implies that inside the linearly unstable region,
there exists a range of density where the homogeneous flow is 
{\it convectively} stable \cite{Manneville}
(that is, the instability drifts away in one particular
direction leaving the regions behind unaffected).
The same observation is made in Ref.~\cite{Helbingcondmat},
where the HCT state is related to the stationary synchronized flow.
As mentioned in the preceding subsection, however,
the density-flow relation in congested region of the HCT state lies on 
the single curve $(\rho,\rho V(\rho))$, which differs from the 
experimental observations of scattered data \cite{Kerner1}.
We mention that the HCT state may be very sensitive 
to the presence of noises since it is only convectively stable.

 It is instructive to compare the stability ranges 
in Fig.~\ref{fig:phase}(c) with those in Fig.~\ref{fig:phase}(b).
The stability range of the SLC state lies entirely
below $f_{\rm rmp}^{\rm b}$ in Fig.~\ref{fig:phase}(c) 
while it is mostly above $f_{\rm rmp}^{\rm b}$
in Fig.~\ref{fig:phase}(b).
Also the stability range of the OCT reaches below $f_{\rm rmp}^{\rm b}$
in Fig.~\ref{fig:phase}(c) while it lies entirely
above $f_{\rm rmp}^{\rm b}$ in Fig.~\ref{fig:phase}(b).
This suggests that when the sum
$f_{\rm up}+f_{\rm rmp}$ is the same,
the formation of the cluster is easier for larger $f_{\rm rmp}$
and thus the actual
total flux level in the downstream is lower. 
Physically this tendency can be understood as resulting from the larger density
gradient near the on-ramp
when the relative portion of $f_{\rm rmp}$ is larger.
This trend is indeed observed in highway measurements \cite{Hall}
and is called the ``capacity reduction".

\section{Analytic solutions of the SLC and the HCT states}
In Sec. II, we showed that various forms of traffic flows occur
near an on-ramp. In the case of the free flow and the usual traffic jam,
they are affected by the on-ramp in 
minor ways and their properties are essentially the same
as those without an on-ramp, which have already been
investigated  intensively \cite{Kerner4}.
For other phases, however, the presence of the on-ramp is
crucial and understanding of their properties are 
relatively poor.
In this Section, we present analytic studies of two forms
of traffic flow, the SLC state
and the HCT state. 

\subsection{Standing localized cluster (SLC) state}
The analytic examination of the SLC state is relatively
simple since all time dependence disappears.
This analysis also provides a good starting point
for future analysis of the RH state and the OCT
state since they are closely related to the SLC
state as discussed in the preceding Section.
Hence we present below the analysis of the SLC state in detail.

In a homogeneous highway, inhomogeneities in the density
or the velocity always propagate. In the presence of
an on-ramp, on the other hand, the numerical investigation
in the previous Section shows that inhomogeneities may 
form a {\it standing} cluster without propagation.
Here we demonstrate analytically that 
the model [Eqs.~(\ref{eq:eqofmotion1},\ref{eq:eqofmotion2})] 
indeed allows standing cluster solutions
in the presence of an on-ramp.

To obtain the SLC solution, one imposes
\begin{equation}
{\partial \rho \over \partial t}={\partial v \over \partial t}=0 \ .
\label{eq:stationary}
\end{equation}
By integrating Eq.~(1) with respect to $x$,
one obtains
\begin{equation}
\rho(x) v(x)=f_{\rm rmp} \int_{-\infty}^x \varphi (x) dx
+f_{\rm up} \equiv q(x) \ .
\label{eq:integrating}
\end{equation}
Since the function $q(x)$ is completely determined for given 
$f_{\rm rmp}$ and $f_{\rm up}$,
one can use this equation to express $\rho(x)$ 
in terms of $v(x)$.
Using this, one can rewrite Eq.~(\ref{eq:eqofmotion2})
as follows:
\begin{equation}
\mu {d^2 v \over d x^2}=q \left(1-{c_0^2 \over v^2}\right) {d v \over dx}
-{q(x) \over \tau v} \Biggl[V \left( {q(x) \over v} \right)-v\Biggr]
+{c_0^2 \over v} q'(x) \ ,
\label{eq:slcequation}
\end{equation}
where $\partial/\partial x$ has been replaced by $d/dx$
since all time dependence disappears. 

 For further analysis, it is convenient to assume
a particular form of the influx profile $\varphi (x)$.
We take the localized influx limit and choose $\varphi (x)=\delta (x)$.
Then $q(x)$ becomes $f_{\rm up}$ for $x<0$, $f_{\rm up}+f_{\rm rmp}$
for $x>0$, and $q'(x)=f_{\rm rmp}\delta (x)$.
Then Eq.~(\ref{eq:slcequation}) can be decomposed
into two separate problems
defined on two semi-infinite regions,
$x<0$ and $x>0$, with the matching conditions at $x=0$ \ ,
\begin{eqnarray}
v(x)|_{x=0+}&=&v(x)|_{x=0-} \ , \nonumber \\
\left. {dv\over dx} \right|_{x=0+}&=&
\left. {dv\over dx} \right|_{x=0-}+{c_0^2 \over \mu v(0)}f_{\rm rmp} \ . 
\label{eq:BC}
\end{eqnarray}

 For each semi-infinite region, it
is instructive to rewrite Eq.~(\ref{eq:slcequation})
as follows,
\begin{eqnarray}
\mu {dw \over dx}&=&q_{\rm s}\left(1-{c_0^2\over v^2}\right)w-
 {q_{\rm s} \over \tau v}\left[ V\left({q_{\rm s} \over v}\right)-v \right] \ ,
  \nonumber \\
{dv \over dx}&=& w \ , \label{eq:evolution} 
\end{eqnarray}
where s=p(ositive) for $x>0$ and s=n(egative) for $x<0$,
and $q_{\rm p}=f_{\rm up}+f_{\rm rmp}$, $q_{\rm n}=f_{\rm up}$.
Notice that after the variable transformations $v \rightarrow y,
x \rightarrow t, \mu \rightarrow m$,
Eq.~(\ref{eq:evolution}) can be regarded as the equation
of motion of a particle
subject to a potential $U_{\rm s}(y)$ where
${d U_{\rm s} / d y}=(q_{\rm s} / {\tau y})[V({q_{\rm s}/ y})-y]$
and to the ``strange'' coordinate-dependent damping force.

 For the safe velocity $V(\rho)$ adopted in this paper,
Eq.~(\ref{eq:evolution}) is highly nonlinear and it
does not seem feasible to write down solutions in
a closed form.
However, qualitative properties of the solutions
can be still investigated by taking Eq.~(\ref{eq:evolution})
as a set of flow equations defined on the phase space $(v,w)$.

Since the global structure of the flow is largely determined
from properties of fixed points,
we first find fixed points of Eq.~(\ref{eq:evolution}).
Simple algebra shows that there are
three fixed points,
$(v,w)=(0,0), (v_{\rm s1},0), (v_{\rm s2},0)$.
The first one is unphysical since $v=0$ implies
$\rho \rightarrow \infty$.
This unphysical fixed point appears since we set $V(\rho)=0$
for $\rho > {\hat \rho}$, and we ignore this below.
The other two come from the two solutions $v_{\rm s1}$,
$v_{\rm s2}$ $( < v_{\rm s1})$ of $V(q_{\rm s}/v)=v$.
(It can be easily verified that for $q_{\rm s} <f_{\rm max}$,
there are always two solutions, the larger one
corresponding to the maximum point 
of the potential $U_{\rm s}(y)$ and the smaller to the
minimum point.)

\begin{figure}
\begin{center}
\epsfig{file=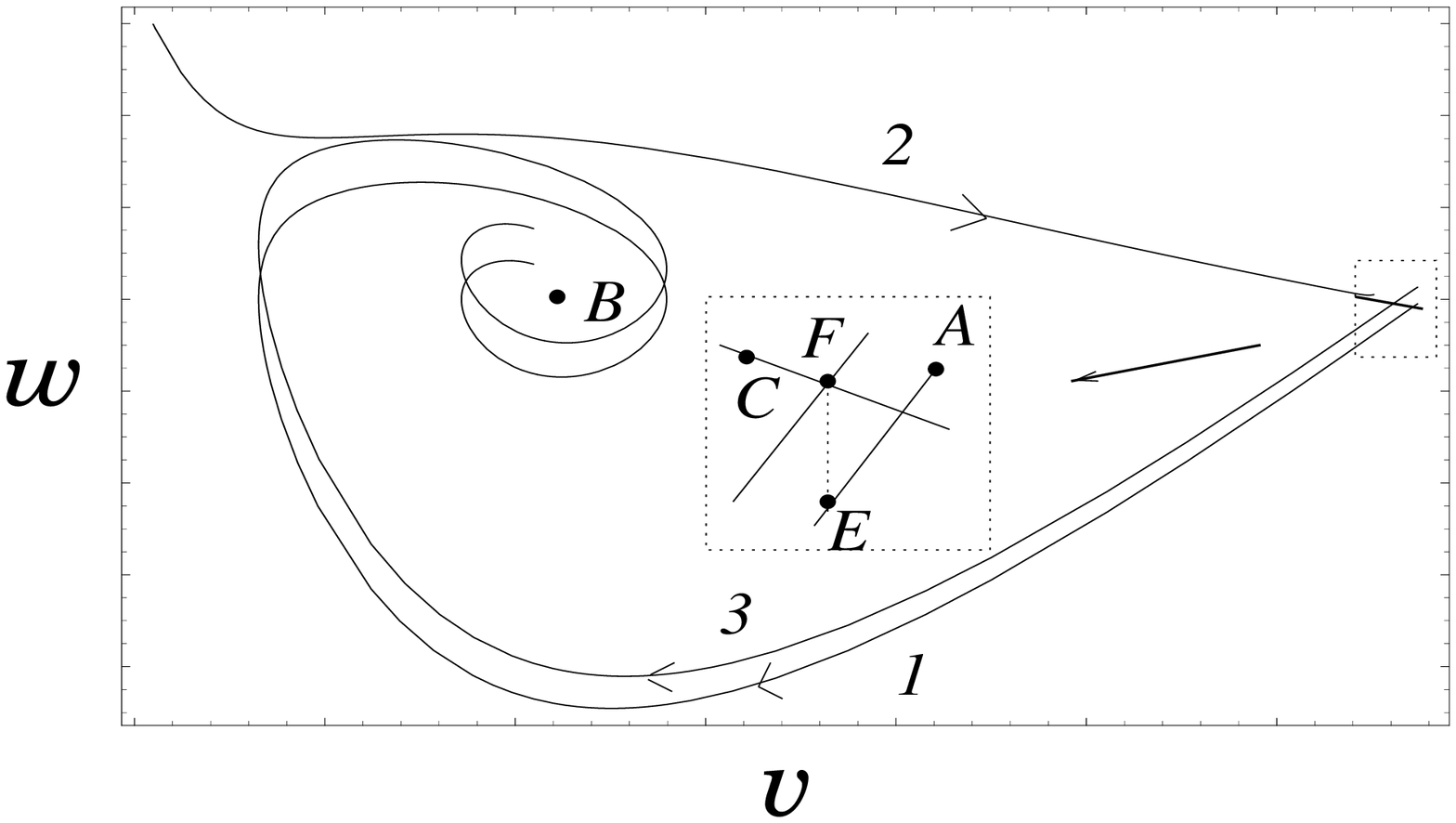,clip=,width=0.95\columnwidth}(a)\\
\epsfig{file=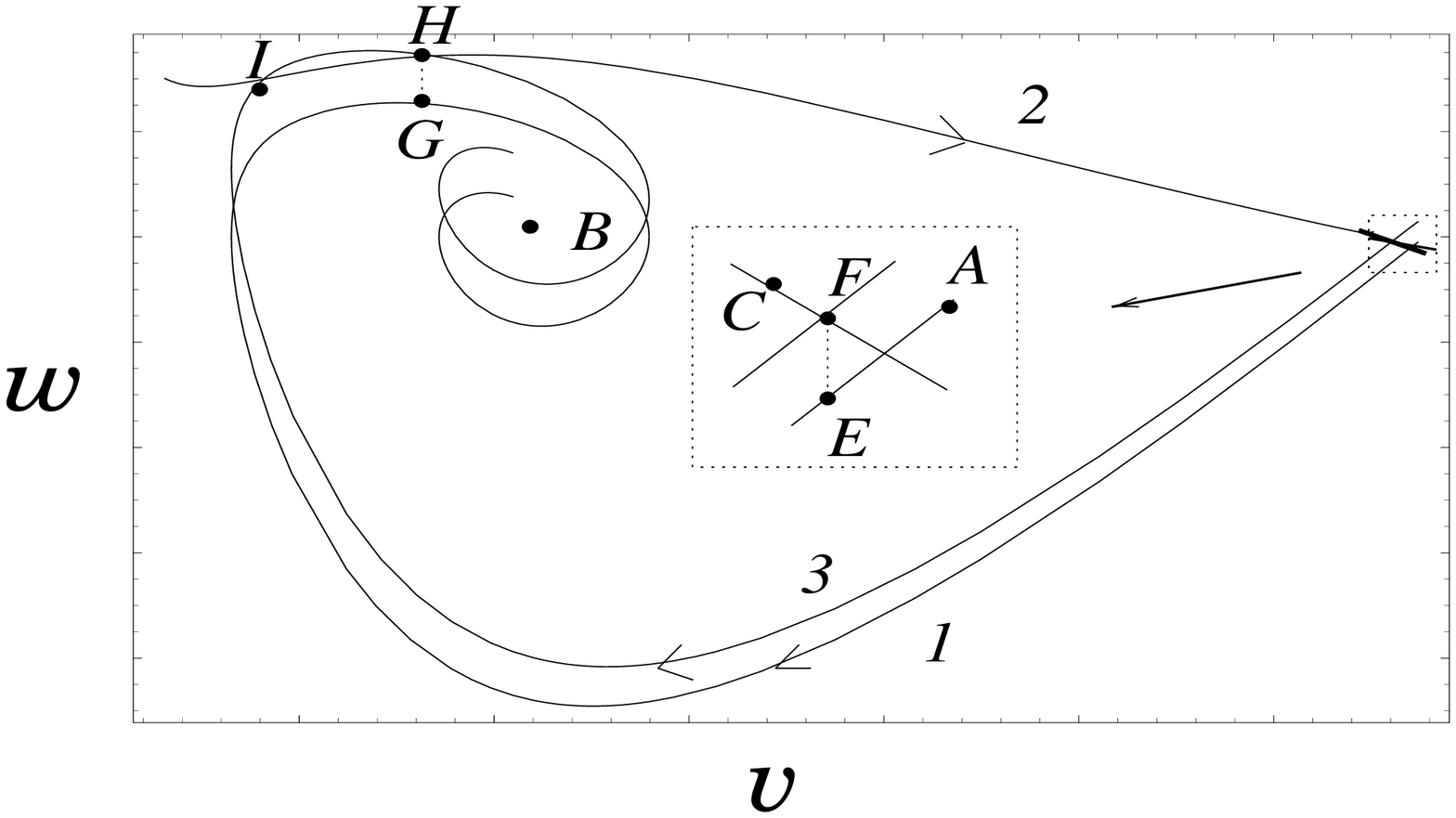,clip=,width=0.95\columnwidth}(b)\\
\epsfig{file=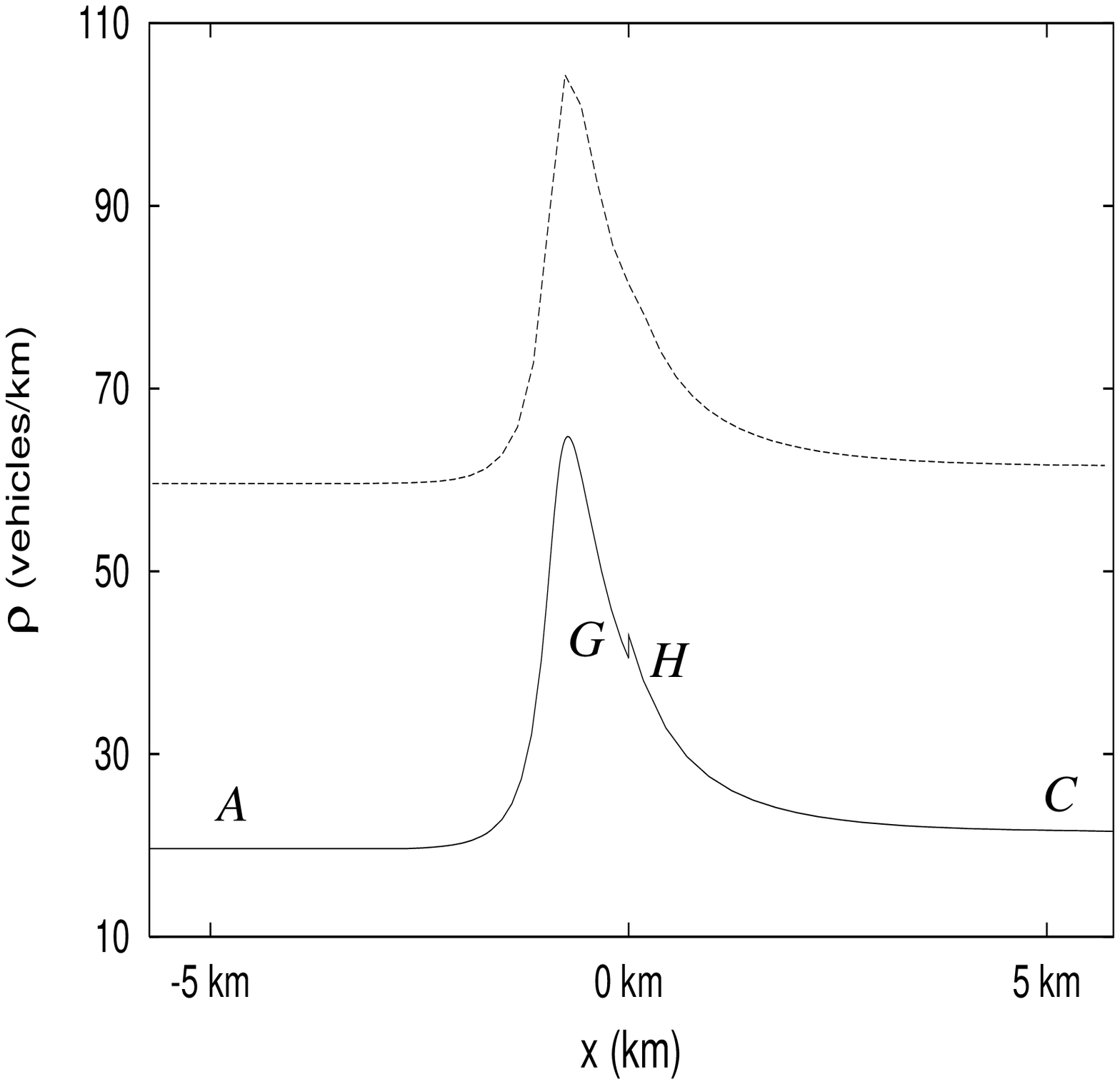,clip=,width=0.95\columnwidth}(c)\\
\end{center}
\caption[]{
The flow in the phase space $(v,w)$
(a) with $f_{\rm rmp}$ slightly below the lower stability limit
of the SLC state and (b) with $f_{\rm rmp}$ slightly above
the lower stability limit. 
The path 1 represents the flow
that is associated with the unstable eigen-direction
of the fixed point $(v_{\rm n1},0)$ (point $A$),
and the path 2 the flow that is associated with
the stable eigen-direction of the fixed point $(v_{rm p1},0)$
(point $C$).
The path 3 represents the image of the path 1 through
the mapping (\ref{eq:BC}).
The insets are the enlargements of the dotted regions
at right.
(c) The density profile (solid line) of the SLC state
obtained analytically for $f_{\rm up}$=1948 vehicles/h
and $f_{\rm rmp}$=121 vehicles/h.
The marks $A$, $G$, $H$, $C$
represent 
corresponding points in Fig.~\ref{fig:flowdiagram}(b).
The density profile 
obtained from the numerical simulation of Eqs.~(\ref{eq:eqofmotion1},
\ref{eq:eqofmotion2}) is also given for comparison
(dotted line, vertically shifted 40 vehicles/km). 
Notice that the density jump at the on-ramp ($x=0$)
in the solid line (due to the approximation $\varphi(x)=\delta (x)$)
is smoothed out in the dotted line.
}
\label{fig:flowdiagram}
\end{figure}

We are interested in the solutions where
\begin{equation}
v(x) \rightarrow  \left \{ \begin{array}{ll} v_{\rm n1}&~~ \mbox{for}~~ x 
                           \rightarrow -\infty \nonumber \\
                      v_{\rm p1}&~~ \mbox{for}~~ x \rightarrow \infty . 
                       \end{array} \right .
\end{equation}
Thus $(v_{\rm n1},0)$ [point $A$ in Fig.~\ref{fig:flowdiagram}(a,b)]
is the relevant fixed point
for $x<0$. By linearizing Eq.~(\ref{eq:evolution}),
it can be verified that it is a saddle point.
Then the flow [path 1 in Fig.~\ref{fig:flowdiagram}(a,b)]
that is associated with the unstable eigen-direction
of the fixed point determines the entire flow in 
the semi-infinite region $x<0$.
Similarly $(v_{\rm p1},0)$ [point $C$ in Fig.~\ref{fig:flowdiagram}(a,b)]
is the relevant fixed point
for $x>0$, which is again a saddle point.
The entire flow in the positive semi-infinite region
is then determined by the flow [path 2 in Fig.~\ref{fig:flowdiagram}(a,b)]
that is associated with
the stable eigen-direction
of the fixed point.
 
 In order to construct a legitimate solution
from the path 1 and 2, one should join the two paths using the matching
conditions [Eq.~(\ref{eq:BC})].
It is convenient to regard the conditions [Eq.~(\ref{eq:BC})]
as a definition of a mapping defined in the phase
space $(v,w)$,
from a point $(v,w)$ to $(v,w+c_0^2 f_{\rm rmp}/ \mu v)$.
The effect of the mapping is shown in Fig.~\ref{fig:flowdiagram}(a),
where the path 1 is mapped to the path 3.
The path 3 crosses the path 2 at a point denoted as $F$
in the inset.
Then one can construct a full solution
by connecting the curve $AE$ with the curve $FC$.
This solution exists for
an arbitrary small value of $f_{\rm rmp}$
and represents the free flow solution (with a transition
layer \cite{Kerner95} at the on-ramp).

 Different solutions appear for sufficiently large $f_{\rm rmp}$.
The mapping of the path 1 to the path 3 for a larger $f_{\rm rmp}$ 
is depicted in Fig.~\ref{fig:flowdiagram}(b).
Now the path 3 crosses the path 2 at three points
$F, H, I$, which implies the presence of three solutions.
The solution associated with 
the crossing point $F$ again
corresponds to the transition layer solution.
The two other solutions associated with the crossing points $H$ and $I$ 
correspond to the desired SLC solutions.
Each solution provides the velocity profile for $-\infty < x < \infty$,
from which the density profile can be obtained from $\rho(x)=
q_{\rm s}/v(x)$.
Fig.~\ref{fig:flowdiagram}(c) compares the
density profile associated with the crossing point $H$ with that from 
the direct numerical simulation of the traffic model 
[Eqs.~(\ref{eq:eqofmotion1},\ref{eq:eqofmotion2})]
with the Gaussian form of $\varphi(x)$. 
Notice that the two profiles are essentially identical except
for a small difference in the ramp region,
which arises due to the approximation of the input flux
profile by a delta function.
The density profile associated with the crossing
point $I$ can be obtained in a similar way.
This solution, however, is not found in the direct numerical
simulation,
which suggests that this solution is linearly unstable.

 This analysis implies that the SLC solutions appear only for
$f_{\rm rmp}$ larger than a critical value.
Precisely at the critical value, the two crossing points
$H$ and $I$ coincide.
For $f_{\rm rmp}$ larger than the critical value,
numerical simulations indicate that only
one solution
(one through the crossing point $H$) is stable
and the other is not.
Thus one finds that a turning point connecting the
stable and unstable SLC solutions appear
at the critical ramp flux.

Above considerations use a specific form of an influx profile.
We believe that the precise profile does not
change the qualitative nature of above discussion.
In fact, the value of the critical ramp flux for the SLC state
obtained using the approximation $\varphi(x)=\delta(x)$ is
found identical within the numerical accuracy 
to that obtained from the direct numerical
simulation of Eqs.~(\ref{eq:eqofmotion1},\ref{eq:eqofmotion2})
(with $\varphi(x)$ Gaussian).
%

\subsection{Homogeneous congested traffic (HCT) state}

When the on-ramp influx is added, we
observe a new kind of the traffic jam:
The downstream front is fixed at the on-ramp
and the upstream front moves with a fixed group velocity.
Between the downstream and the upstream fronts,
the congested region of the homogeneous flow 
is maintained [Fig.~\ref{fig:OCTstructure}(b)].
We notice that the upstream front is a steady structure
in a proper reference frame.
Below we show analytically that Eqs.~[\ref{eq:eqofmotion1},
\ref{eq:eqofmotion2}] possess the HCT state solution.
Since the congested region is homogeneous,
one can split the discussion into two parts,
one for the moving upstream front and the other for the fixed
downstream front.
For the upstream front,
the relevant equations of motion are
\begin{eqnarray}
{\partial \rho \over \partial t}+{\partial (\rho v) \over \partial x}
&=&0 \ , \label{eq:heqofmotion1} \\ 
\rho \left({\partial v \over \partial t}+v{\partial v \over \partial x}\right)
&=&{\rho \over \tau}[V(\rho)-v] -c^2_0 \ {\partial \rho \over \partial x}
+\mu {\partial^2 v \over \partial x^2} \ .
\label{eq:heqofmotion2}
\end{eqnarray}
Since this front is surrounded by
{\it wide} regions of the homogeneous flow
both in the upstream and the downstream,
one can impose the following boundary conditions,
\begin{eqnarray}
\rho(x=-\infty, t)&=&\rho_{-}=\rho_{\rm up} \ , \label{eq:hbc1} \\
v(x=-\infty,t)&=&v_{-} = V(\rho_{-}) \ , \label{eq:hbc2}\\
\rho(x=+\infty, t)&=&\rho_{+} \ , \label{eq:hbc3}\\ 
v(x=+\infty,t)&=&v_{+} = V(\rho_{+}) \ .
\label{eq:hbc4} 
\end{eqnarray}
Here $\rho_{-}$ is the density of the far upstream region
and $\rho_{+}$ of the congested region 
Also the spatial coordinate is chosen in such a way that
$x=+\infty$ corresponds to a location deep in the
congested region instead of the far downstream in the 
original equations [\ref{eq:eqofmotion1},\ref{eq:eqofmotion2}]. 

Let us assume that Eqs.~(\ref{eq:heqofmotion1}) and (\ref{eq:heqofmotion2})
allow a {\it steady} state solution that satisfies the 
boundary conditions Eqs.~(\ref{eq:hbc1}), (\ref{eq:hbc2}), (\ref{eq:hbc3}),
and (\ref{eq:hbc4}).
Here the steady state means that in a proper reference frame,
all time-dependence disappears.
We perform the change of the reference frame:
\begin{eqnarray}
x'&=&x-v_g t \ , \\
t'&=&t \ ,
\end{eqnarray}
and neglect all time-dependence in this new reference frame
to find
\begin{eqnarray}
{d q' \over d x'}&=&0 \ , \label{eq:hpem1} \\
q' {d v \over d x'}&=&\rho {V(\rho)-v \over \tau} -c_0^2 {d \rho \over d x'}
+\mu {d^2 v \over d x'^2} \ , \label{eq:hpem2} 
\end{eqnarray}
where $q'\equiv \rho v- \rho v_g$.
 We can determine the constants $q'$ and $v_g$ 
from the boundary conditions Eqs.~(\ref{eq:hbc1}), (\ref{eq:hbc2}),
(\ref{eq:hbc3}), and (\ref{eq:hbc4}).
From the condition that the in-flux to
and the out-flux from the front 
should be the same in the primed reference frame,
we obtain
\begin{eqnarray}
v_g&=&{{\rho_{+} v_{+}-\rho_{-} v_{-}}
\over {\rho_{+}-\rho_{-}}} \ , \label{eq:vg}\\
q'&=& {{\rho_{+}\rho_{-}(v_{-}-v_{+})}
\over {\rho_{+}-\rho_{-}}} \ .
\end{eqnarray}

Using Eq.~(\ref{eq:hpem1}), one has
\begin{equation}
\rho={q' \over v-v_g} \ .
\end{equation}
Plugging in this expression into Eq.~(\ref{eq:hpem2}), one finds
\begin{equation}
\mu {d^2 v \over d x'^2}+q'\left[{c_0^2 \over (v-v_g)^2}-1\right] 
{d v \over d x'}
+F(v;q',v_g)=0 \ , \label{eq:eqofmass}
\end{equation}
where 
\begin{equation}
F(v;q',v_g) \equiv {q' \over \tau (v-v_g)}
\left[V\left ({q' \over {v-v_g}} \right )-v \right] \ .
\end{equation}
Eq.~(\ref{eq:eqofmass}) is again
an equation of motion of a particle subject
to a conservative force $-F$ and the
coordinate-dependent damping force.
The boundary conditions Eqs.~(\ref{eq:hbc1}), (\ref{eq:hbc2})
ensure that 
$F(v_{\pm };q',v_g)$=0 automatically. 
Since $v_{-} > v_{+}$,
the root $v_{-} (v_{+})$ corresponds to the potential maximum (minimum).
Thus the nature of the stationary
state is clear in the
particle motion analogy.
At $x=-\infty$, the particle is at the unstable
maximum point $v=v_{-}$.
As ``time'' $x$ increases, the particle
slides down the hill and after some time,
it settles down at $v=v_+$ due to the friction,
provided the damping coefficient in Eq.~(\ref{eq:eqofmass})
remains positive.
This particle motion describes the upstream
front of the HCT state [Fig.~\ref{fig:OCTstructure}(b)].

 Before we begin the next analysis of the downstream
front of the HCT state [Fig.~\ref{fig:OCTstructure}(b)],
one remark is in order.
In the direct numerical simulation of Eqs.~[\ref{eq:eqofmotion1},
\ref{eq:eqofmotion2}],
the vehicle velocity in the congested region of the HCT state
is fixed for given $f_{\rm up}$ and $f_{\rm rmp}$.
However, in the above analysis of the upstream front,
where $f_{\rm up}$ is used to fix $v_{-}=v_{\rm up}$,
the value of $v_{+}$ is still a free parameter.
We show below that this degree of freedom should be used to
allow the downstream front solution.

 Analysis of the stationary downstream front is very similar
to the analysis of the SLC state in the
preceding subsection.
Using the condition of the stationarity and 
integrating Eq.~(\ref{eq:eqofmotion1}),
one obtains Eqs.~(\ref{eq:stationary}) and (\ref{eq:integrating}), 
respectively.
Adopting the approximation $\varphi (x)=\delta (x)$,
one recovers the matching conditions
at the ramp [Eq.~(\ref{eq:BC})] and
the set of the flow equations [Eq.~(\ref{eq:evolution})].
The value of $q_{\rm n}$ should be $\rho_+ v_+$
to allow a continuous connection of $v(x)$ to the upstream
front solution and $q_{\rm p}=q_{\rm n}+f_{\rm rmp}$.
The asymptotic behavior of $v(x)$ far away from the on-ramp
should be chosen differently as follows:
\begin{equation}
v(x) \rightarrow  \left \{ \begin{array}{ll} 
  v_{\rm n2}=v_+&~~ \mbox{for}~~ x \rightarrow -\infty \  \nonumber \\
  v_{\rm p1}&~~ \mbox{for}~~ x \rightarrow \infty \ ,                      
 \end{array} \right .
\end{equation}
where $v_{\rm p1}$ and $v_{\rm n2}$ are defined
in the same way as in the preceding subsection,
and $x \rightarrow -\infty$ corresponds to the region deep in the congested
region.

Notice that for $x \rightarrow -\infty$,
$v(x)$ approaches $v_{\rm n2}$ instead of $v_{\rm n1}$
since $v_+$ corresponds to the smaller of the two solutions of
$V(q_{\rm n}/v)=v$ [or since the corresponding
density $\rho_+$ lies on the descending slope of the homogeneous
density-flow relation]. 
This difference in the asymptotic behavior results in a qualitative
change. It can be verified through the linearization
of Eq.~(\ref{eq:evolution}) that 
$(v_{\rm n2},0)$ is a {\it stable} fixed point.
Then we find $v(x)=v_{\rm n2}$ for the entire semi-infinite
region $x < 0$, which should be contrasted to
the preceding subsection.
Near the fixed point $(v_{\rm p1},0)$,
on the other hand, the situation is similar to 
the preceding subsection and
the flow in the positive semi-infinite region becomes
a continuous path, like the path 2 in Fig.~\ref{fig:flowdiagram}(a,b).

 To construct a full solution of the HCT downstream front,
the separate solutions for $x<0$ and $x>0$ should be
joined using the matching condition [Eq.~(\ref{eq:BC})].
Since the velocity for $x<0$ is constant,
the matching conditions reduce to
\begin{eqnarray}
v(x)|_{x=0+}&=&v_{\rm n2}, \nonumber \\
\left. {dv\over dx} \right|_{x=0+}&=&
{c_0^2 \over \mu v(0)}f_{\rm rmp} \ . 
\label{eq:HCTBC}
\end{eqnarray}
This matching conditions can be satisfied only
when the phase space trajectory for $x>0$
passes the point 
$(v_{n2}, {c_0^2 / (\mu v_{\rm n2})}f_{\rm rmp})$.
For given $v_{\rm n2}$ and $f_{\rm rmp}$,
the path in general does {\it not} pass the point 
except for a particular value of $v_{\rm n2}=v_+$.
This tuning thus fixes the free parameter $v_+$ as mentioned before.
In general, $v_+$ is a function of the influx profile $\varphi(x)$
since the matching conditions depend on $\varphi (x)$.
 
\begin{figure}
\begin{center}
\epsfig{file=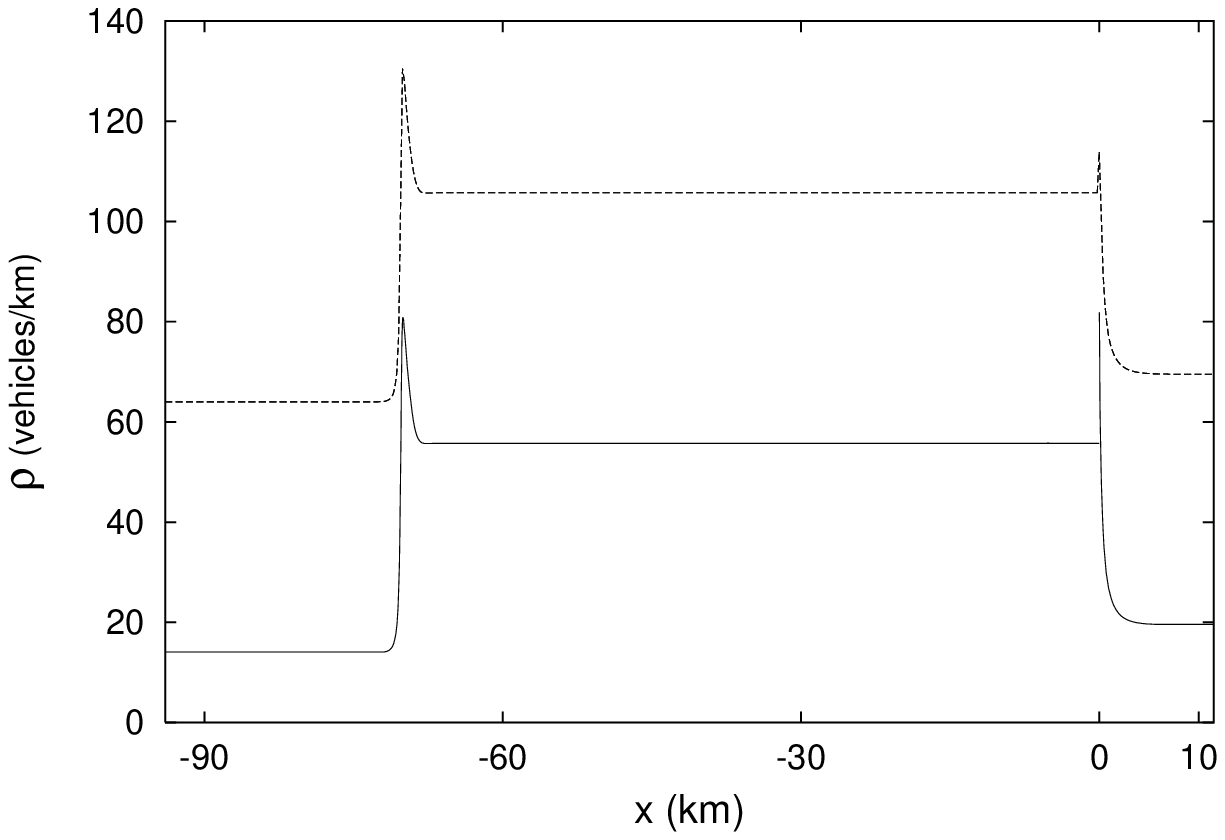,clip=,width=0.95\columnwidth}
\end{center}
\caption[]{
The density profile of the HCT state
obtained analytically (solid line) under
the approximation $\varphi(x)=\delta (x)$
for $f_{\rm up}=1497$ vehicles/h
and $f_{\rm rmp}$=762 vehicles/h.
The density profile 
obtained from the numerical simulation of 
Eqs.~(\ref{eq:eqofmotion1},\ref{eq:eqofmotion2}) is also given for comparison
(dotted line, vertically shifted 50 vehicles/km).
Small differences between the two are due to 
the different choices of $\varphi(x)$.
}
\label{fig:analhct}
\end{figure}

 Fig.~\ref{fig:analhct} shows the density profile (solid line)
of the HCT state obtained from this matching 
method for $\varphi (x) =\delta (x)$.
It is essentially identical to the profile (dotted line) obtained
from the numerical simulation of 
Eqs.~(\ref{eq:eqofmotion1},\ref{eq:eqofmotion2}), except for the larger density peak at the 
on-ramp caused by the approximation $\varphi(x)=\delta(x)$.
We confirm from the simulation result that the group velocity 
of the moving front 
is consistent with $-7.68$ km/h which
is given by Eq.~(\ref{eq:vg})
and the damping coefficient $q'[{c_0^2 / (v-v_g)^2}-1]$ is 
positive in the congested region.

\section{CONCLUSION}
The traffic equation with a source term representing
the on-ramp influx
of a highway displays a variety of novel traffic flow states
not present in the homogeneous equations.
To understand the role of the source term,
we map out the phase diagram 
using the continuum traffic model 
[Eqs.~(\ref{eq:eqofmotion1},\ref{eq:eqofmotion2})]
proposed by Kerner and Konh\"{a}user \cite{Kerner5}.
In our numerical simulation, we use the open boundary condition,
which allows one to handle the single on-ramp
without using very large system sizes.
Due to possible presence of multiple metastable states,
detailed simulation is carried out for a limited number of representative
values of the upstream flux $f_{\rm up}$ 
and for the whole range of the on-ramp flux $f_{\rm rmp}$.
Various traffic states are identified and characterized.
The phase diagrams thus obtained are summarized in Fig.~\ref{fig:phase}.
It is found that an inhomogeneous but stationary
traffic state (SLC) can appear near the on-ramp due to $f_{\rm rmp}$.
This state is related to the recent measurement of the homogeneous
synchronized flow.
The on-ramp also generates a new kind of traffic jam,
which can appear even below the stability limit of the usual traffic
jam in homogeneous highways.
The structure of the new traffic jam varies qualitatively
with $f_{\rm rmp}$.
The capacity reduction due to $f_{\rm rmp}$ is also observed.
In a certain range of $f_{\rm rmp}$, the free flow, the RH state,
and the new traffic jam can all coexist so that
the free flow can undergo phase transitions either to
the RH state or to the traffic jam state.
Analytic investigations are also performed and two 
nontrivial solutions of Eqs.~(\ref{eq:eqofmotion1},\ref{eq:eqofmotion2})
are found. These solutions describe the SLC state and the
HCT state.

\section*{ACKNOWLEDGMENTS}
H.Y.L. thanks Daewoo Foundation for financial support,
and M. Schreckenberg for hospitality during her stay 
at Duisburg University.
H.-W.L. is supported by the Korea Science and Engineering Foundation
through a fellowship.
This work is supported by the Korea Science and Engineering
Foundation through the SRC program at SNU-CTP,
and also by Korea Research Foundation (1998-015-D00055).

\end{multicols}


\begin{thebibliography}{7}
\bibitem{commentV} The form of $V(\rho)$ is chosen so as to be
  consistent with the available free flow 
  data~\protect\cite{Kernerl} and also with large density
  asymptotic behaviors. But its precise form is not important.
\end{thebibliography}
\end{document}